\documentclass[aps,pra,twocolumn,superscriptaddress]{revtex4-2}
\usepackage{amsmath,amssymb,amsfonts,bbm,graphicx,times,psfrag}
\usepackage[pdftex]{color}
\usepackage{amsmath,bm}
\usepackage[colorlinks, linkcolor=blue, citecolor=red, urlcolor=red, breaklinks=true]{hyperref}
\usepackage{microtype}

\usepackage{amsthm}
\usepackage{csquotes}
\usepackage{amsthm}\newcommand{\be} {\begin{equation}}
\newcommand{\ee} {\end{equation}}

\usepackage{color}
\usepackage{mathrsfs}
\usepackage [english]{babel}
\usepackage{csquotes}
\newcommand{\bla}{bla\\bla\\bla\\bla\\bla}
\newcommand{\colr}[1]{{\leavevmode\color{red} #1}}

\usepackage{filecontents}
\begin{document}
 \title{Is there charged dark matter bound to ordinary matter?  Can it produce observable quantum effects?}
\author{Muhammad Asjad}
 \affiliation{Mathematics Department, Khalifa University of Science and Technology, 127788, Abu Dhabi, United Arab Emirates}
 \author{Paolo Tombesi}
 \affiliation{School of Science and Technology, Physics Division, University of Camerino, 62032 Camerino (MC), Italy}
\begin{abstract}
Levitated nano-spheres of silica, optically trapped in a Fabry-Perot cavity with a single trapping field and the electrostatic field of a charged ring electrode, are used to infer the potential existence of dark matter particles with infinitesimal charge. These particles are presumed to exist in bulk matter as relics of the primordial Universe. In the absence of infinitesimally charged particles within the chosen nano-sphere, the output light in this setup should be thermal. However, if these particles do exist, the cavity's output light is expected to be squeezed even at room temperature, and one could observe entanglement between light and the nano-sphere's center of mass.
\end{abstract}
  \date{\today}
\maketitle
%\paragraph*{Introduction:}
 Dark matter, whose nature remains unclear, is inferred from astrophysical and cosmological observations due to gravitational effects \cite{zwicky, rubin, RubinA, Bertone}. Unlike visible matter, dark matter does not interact with any known force, except through gravity. However, there is a possibility of a photon being directly emitted by dark matter particles, if some of them possessed an extremely weak electric charge difficult to observe but distinguishable through very sensitive measurements \cite{holdom}.

The concept of mini-charged or milli-charged particles (mCPs) allows for fractional or infinitesimal charges beyond traditional quantization \cite{holdom, holdomA, nath, glashow, batell, dutra, jeackel}. These particles carry charges represented as $q=\pm\epsilon |e_0|$, where $\epsilon$ is estimated to range from $10^{-16}$ to $10^{-2}$, and $e_0$ is the charge on an electron \cite{Ahlers}. mCPs might also be bound to ordinary matter nuclei, leading to milli-charged atoms \cite{holdomA, steigman, cline}. Although their interaction with electromagnetic fields is extremely weak, it cannot be excluded. Numerous experiments have sought these elusive particles using colliders and levitated particles in bulk matter or as isolated particles \cite{steigman, morpurgo, morpurgoA, gratta, zurek}, placing limits on their abundance for certain values of $\epsilon$ \cite{afek}. Recently, a direct search using the PandaX-4T xenon-based detector has provided a charge limit $< 2.6\times 10^{-11} e_0$ for an estimated dark matter mass of $20-40$ $\rm{G e V/c}^2$ by investigating effective electromagnetic interactions between dark matter particles and xenon nuclei \cite{panda}.

Trapping and levitation of microscopic dielectric spheres have been studied since the 1970s \cite{ashkina,ashkinb}. Stable trapped droplets have been maintained at pressures as low as $10^{-6}$ Torr \cite{ashkinc}. Levitating dielectric particles in optical cavities is now a field of interest in quantum optomechanics, enabling the creation of an oscillating massive system nearly decoupled from the environment. This opens up possibilities for cooling mesoscopic objects to their quantum ground state and developing highly sensitive measurement devices.

This letter discusses the search for mini-charged particles (mCPs) in bulk matter using optically trapped fused silica ($\text{SiO}_2$) nano-spheres (NS). The levitated NS here considered  has a radius $r= 50$ nm, significantly smaller than the trapping/driving laser's optical wavelength ($\lambda = 1064$ nm) in a single mode ideal one-sided Fabry-Perot cavity. The cavity operates at a frequency of $\omega_{\rm{c}}/2\pi$, enclosed in a controlled-temperature vacuum chamber with air pressure around $10^{-10}$ Torr. It is assumed that the mCPs' mass falls within a range compatible with previous results \cite{davidson}, possibly around or greater than one $\rm{MeV/c}^2$.

Ensuring no induced electron charge on the NS surface during its generation is crucial \cite{gratta}. However, it is plausible that the silica droplet may not contain any mCPs, given the unknown abundance of these charged dark matter relics in terrestrial materials \cite{afek}. Therefore, several experiments with different silica droplets to obtain a mass of at least one mg would be necessary before drawing some conclusions about mCPs' potential existence \cite{supmatb}. With the experimental set-up considered here, if mCPs do exist in the silica droplet, the output light deviates from pure thermal noise, revealing squeezing in the measured symmetric spectrum of light's quadrature fluctuations. Another significant quantum effect that could be observed is the entanglement between light and the center of mass (CoM) of the NS.

%\paragraph*{System and Equation of motions:}
The assumption is that the NS is trapped at one antinode of the steady field within a one-sided ideal Fabry-Perot cavity, positioned at the centre, far from the cavity mirrors, to eliminate any contamination from the Casimir force effect \cite{nie}. Previous studies have demonstrated that the oscillatory motions of the NS along the three axes are mostly independent of each other \cite{novotny}. Therefore, in this analysis, it will be considered only the motion along the cavity axis, assuming that the motion along the other two directions is confined to a region much smaller than the waist of the trapping field. This condition can be achieved, for instance, by implementing feedback cooling methods to confine the transverse motions \cite{raizen}. In recent years, several authors have studied such a system (e.g., Refs. \cite{kimble, isart, aspelmeyer, monteiro, kiesel} ). However, in this discussion, a simplified scheme will be followed, focusing solely on a single trapping mode without considering the cooling mode.

Assuming that there is an mCP trapped within the perfectly transparent silica nano-sphere, which could be bound to an atom of silicon or oxygen, several approaches have been considered to enhance its interaction with our world. These approaches include: (i) positioning the cavity axis between two parallel electrodes connected to highly different voltage sources, as described in Ref. \cite{afek}; (ii) using a strongly charged metallic needle inserted between the cavity mirrors to generate a strong Coulomb interaction with the mCP, as proposed in Ref. \cite{rashid}; and (iii) inserting a uniformly charged metallic ring with a linear charge density \emph{l} and a radius $R>>r$ much larger than the radius of the nano-sphere, with the plane of the ring orthogonal to the cavity axis and centered on it, as outlined in Ref. \cite{zhu}. Although there is also a paper \cite{fonseca} that considers a ``classically charged" microsphere levitated in a hybrid optical-Paul trap, here the proposal of Ref. \cite{zhu} will be followed, which could be considered analogous to the one-dimensional static case discussed in Ref.\cite{fonseca}.
% which could be considered analogous to the one-dimensional static case presented in
Then, the Hamiltonian for the levitated dielectric droplet, considered as point-like \cite{kimble, kiesel}, in a frame rotating at the angular frequency $\omega_{\rm{L}}$ of the driving field is\cite{supmat}:
\begin{eqnarray} \label{newH}
\hat H&=&-\hbar \Delta_{0} \hat{a}^\dagger \hat a  - \hbar g  \cos^2(k x) + \hbar E (\hat a + \hat a^\dagger) \nonumber
\\
&+&\frac{ p^2}{2 m}+q\phi(x).
\end{eqnarray}
The detuning of the cavity field with respect to the driving field is denoted as $\Delta_{0} = \omega_{\rm{L}} - \omega_{\rm{c}}$. The second term in Eq. (\ref{newH})  represents the effect of the dielectric droplet, which slightly modifies the frequency of the cavity mode due to the interaction between the point-like dielectric particle and the cavity field \cite{joannopoulos}, giving rise to the ponderomotive coupling constant $g = 3V_{\rm{s}} (\varepsilon- 1)(\varepsilon + 2)^{-1}\omega_{\rm{c}}/2V_c$ \cite{kimble}, where $\varepsilon= 2.3$ represents the electric permittivity of $\text{SiO}_2$, assumed to be real. $V_{\rm{s}}$ denotes the droplet volume and $V_{\rm{c}} = \pi w^2\mathcal{L}/4$ represents the volume of the cavity field, where $w = \sqrt{{\lambda \mathcal{L}}/{2\pi}}$ is the internal field waist and $\mathcal{L}$ is the cavity length. Here, $\hat{a} (\hat{a}^\dagger)$ represents the boson annihilation (creation) operator of the cavity mode with the commutation relation $[\hat a, \hat a^\dagger] = 1$. The driving field amplitude is denoted as $E = \sqrt{\frac{\kappa \mathcal{P}}{\hbar \omega_{\rm{L}}}}$, where $\kappa = \frac{c \pi}{2\mathcal{L}\mathcal{F}}$ represents the cavity linewidth, $c$ is the speed of light in vacuum, $\mathcal{F}$ is the cavity finesse, and $\mathcal{P}$ is the input power of the trapping laser with frequency $\omega_{\rm{L}}/2\pi$. The mass of the silica droplet is denoted as $m = \rho_0 V_{\rm{s}}$, where $\rho_0 = 2650  \,\mathrm{Kg\, m^{-3}}$ is the mass density. The center of mass position of the NS along the cavity axis is denoted as $x$, and $p$ represents its conjugate momentum with the commutation relation $[x, p] = i\hbar$. Furthermore, $k = {\omega_{\rm{c}}}/{c}$ is $2\pi$ times the inverse wavelength of the trapping field. A polarized Gaussian TEM00 wave along the unit vector $\mathbf{y}$ inside the cavity, of the form $\mathbf{E}(x) = E\mathbf{y}\cos(k x)\, e^{-(y^2+z^2)/w^2}$, has been considered \cite{kimble}. The last term in Eq. (\ref{newH}) represents the electrostatic contribution due to the charged ring, denoted as $q\phi(x)$ where the associated scalar potential $\phi(x)$ is given by
\begin{equation}
\phi(x) = \frac{Q}{4\pi \epsilon_0 R} \left[1 + \left(\frac{C_0 + x}{R}\right)^2\right]^{-\frac{1}{2}},
\end{equation} 
with $Q = 2\pi R l$. 

There are two ways to introduce this charged ring: either by positioning its center at the antinode where the NS's center of mass (CoM) sits in equilibrium, as in Ref. \cite{zhu}, or by placing it at a distance $C_0 \ll \mathcal{L}$ from the trapping antinode on the cavity axis, as considered in this work. Due to its symmetry, the charged ring generates an electrostatic field only along the cavity axis \cite{supmat}, and its effect is incorporated through the introduction of the scalar potential $\phi(x)$. The droplet will be optically trapped at an antinode of the standing wave inside the cavity, and it is convenient to choose this antinode as the origin of the frame, such that $x = 0$ corresponds to the position of its CoM at equilibrium. When $C_0 = 0$, the electrostatic field does not shift the equilibrium position of the NS's CoM; instead, it only modifies its oscillation frequency \cite{zhu}. However, for $C_0 \neq 0$, the equilibrium position $x_{\rm{s}}$ will be shifted away from the anti-node at $x = 0$.

In the presence of strong driving, the Hamiltonian described in Eq. (\ref{newH}) can be linearized by considering the mean steady state values $x_{\rm{s}} (p_s), a_{\rm{s}}$ for the CoM position (momentum) of the NS and the annihilation operator of the cavity mode. This linearization can be expressed as $\mathscr{O} \rightarrow \mathscr{O}_{s} + \delta \mathscr{O}$, where $\mathscr{O}$ represents the operators ($x$, $p$, $\hat a$). Furthermore, in order to ensure that all fluctuation operators are dimensionless a canonical transformation is introduced for the position and momentum fluctuations of the NS's center-of-mass. This transformation is given by $\delta x=\sqrt{\hbar/(m\omega_{\rm{m}})}\delta\hat x$ and $\delta p=\sqrt{\hbar m \omega_{\rm{m}}} \delta\hat p$, where the commutation relation is defined as $[\delta\hat x,\delta\hat p ]= i$. Then, the resulting linearized Heisenberg-Langevin equations of motion, by considering damping and noises, can be obtained \cite{supmat},
\begin{equation} \label{vett}
{\bf{\dot{\hat u}}}(t)={\bf{A}} {\bf{\hat u}}(t)+{\bf{\hat n}}(t),
\end{equation} 
where ${\bf{\hat u}}(t)=(\delta\hat x,\delta\hat p,\delta\hat X,\delta\hat Y)^T$ represents the vector of quadrature operators, and ${\bf{\hat n}}(t)=(0,\hat\eta(t),\sqrt\kappa\hat X_{\rm{in}},\sqrt\kappa\hat Y_{\rm{in}})^T$ corresponds to the noise vector (where the exponent $T$ denotes transposition). In this context, $\delta\hat X =(\delta\hat a +\delta\hat a^\dagger)/\sqrt 2$ and $\delta\hat Y =(\delta\hat a -\delta\hat a^\dagger)/(i\sqrt 2)$ represent the amplitude and phase quadrature fluctuations of the optical field, respectively, while $\hat X_{\rm{in}}=(\hat a_{\rm{in}} +\hat a_{\rm{in}}^\dagger)/\sqrt 2$ and $\hat Y_{\rm{in}}=(\hat a_{\rm{in}} -\hat a_{\rm{in}}^\dagger)/(i\sqrt 2)$ represent the amplitude and phase quadratures of the input noise operators, respectively. Here, $\bf A$ corresponds to the drift matrix:
\begin{equation}\label{matri}
{ \bf A}=\begin{pmatrix}0&\omega_{\rm{m}}&0&0\\-\Omega_{\rm{m}}&-\frac{\gamma}{2}&-G&0\\0&0&-\frac{\kappa}{2}&\Delta(x_{\rm{s}})\\-G&0&-\Delta(x_{\rm{s}})&-\frac{\kappa}{2}\end{pmatrix},
\end{equation} 
where $\Delta(x_{\rm{s}})=\Delta_{0}+g\cos^2(k x_{\rm{s}})$ is the effective detuning, $G=\sqrt{2\hbar/(m \omega_{\rm{m}})}k g a_{\rm{s}} \sin (2kx_{\rm{s}})$ represents the effective coupling and $\Omega_{\rm{m}}=\omega_{\rm{m}}+A_{\rm{q}}/(m\omega_{\rm{m}})$ denotes the effective mechanical frequency with $\omega_{\rm{m}}^2=2 \hbar g k^2 a_{\rm{s}}^2/m\cos(2kx_{\rm{s}})$ and  $A_{\rm{q}}=\frac{q Q}{4\pi\epsilon_0 R^3}$. The decay rate for the cavity mode is denoted by $\kappa$, and for the fluctuation-dissipation theorem \cite{kubo} the damping constant of the mechanical mode, $\gamma$, is given by \cite{supmat}:
\begin{eqnarray}
\gamma &=& \frac{4\pi^2}{5} \frac{\varepsilon-1}{\varepsilon+2} \left(\frac{V_{\rm{s}}}{\lambda^3}\right) \omega_{\rm{m}} \frac{\hbar \omega_{\rm{m}}}{k_{\rm{B}} T} + \frac{4\pi r^2 P_{\text{gas}}}{m v}, \nonumber
\end{eqnarray}
where $P_{\text{gas}}$ denotes the gas pressure, $v=\sqrt{3k_B T/m_{\rm{a}}}$ represents the mean velocity of gas molecules at temperature $T$, $k_{\rm{B}}$ is the Boltzmann constant, and $m_{\rm{a}}=28.97 u $ represents the mass of an air molecule (with $u$ being the atomic mass unit). The stochastic force $\hat\eta(t)$ acting on the mechanical mode which is charchterized by an expectation value of $\langle\hat\eta(t)\rangle=0$ and correlations given by $\frac{1}{2}[\langle\hat\eta(t)\hat\eta(t')\rangle+\langle\hat\eta(t')\hat\eta(t)\rangle]=\Gamma \delta(t-t')$, where $\Gamma= \gamma  k_{\rm{B}} T/(\hbar \omega_{\rm{m}})$ represents the diffusion constant. It is assumed that the stochastic force is Markovian \cite{supmat}.
The optical noise has an expectation value of $\langle\hat J_{\rm{in}}(t)\rangle=0$ and correlations given by $\frac{1}{2}[\langle\hat J_{\rm{in}}(t)\hat J_{\rm{in}}(t')\rangle+\langle\hat J_{\rm{in}}(t')\hat J_{\rm{in}}(t)\rangle]= \frac{1}{2}\delta(t-t')$ (for $\hat J_{\rm{in}}=\hat X_{\rm{in}}(t)$ or $\hat Y_{\rm{in}}(t)$), with the average number of photons at optical frequency being negligible. The steady-state value for optical mode given by $a_{\rm{s}}=E/(i\Delta(x_{\rm{s}})+\kappa/2)$. The phase of input driving laser is chosen such that $a_{\rm{s}}$ is {\it real}, the corresponding steady state value $x_{\rm{s}}$ is obtained by solving the following transcendentalal equation, \cite{supmat} 
\begin{equation} \label{xsvalue}
A_{\rm{q}}(C_0 + x_{\rm{s}})+\frac{\hbar g k E^2 \sin(2kx_{\rm{s}})}{\frac{\kappa^2}{4}+[\Delta_0+g\cos^2(kx_{\rm{s}})]^2}=0
\end{equation}
with the constraints $C_0\not=0$ and $-\frac{\pi}{4k}<x_{\rm{s}}<\frac{\pi}{4k}$. However, unlike typical opto-mechanical systems, the angular frequency $\omega_{\rm{m}} (\Omega_{\rm{m}})$ of the levitated particle, as well as the values of $a_{\rm{s}}$ and $x_{\rm{s}}$, are strongly influenced by the detuning $\Delta_{0}$. Additionally, the value of $\omega_{\rm{m}}$ is also dependent on $\cos(2kx_{\rm{s}})$. Therefore, when choosing $x_{\rm{s}}$, it is imperative to uphold the positivity of the {\it cosine} term in order to preserve the stability of the system \cite{supmat,stb}.

It becomes apparent that the silica droplet and the internal optical field are not independent of each other and the relationship is governed by the {\it non-zero} coupling term $G$, which arises from the milli-charge $q$ and primarily the shift $C_0$ of the charged ring plane. However, this coupling term is case specific, as it disappears when $C_0=0$. Indeed, in this case a different \cite{supmat} approach must be considered because, with the current setting, the only value for light output would be thermal noise alone. This finding already provides an important insight that partially addresses the question posed in the title. With the chosen setup, if a milli-charged particle (mCP) exists within this nano-sphere, the output light should exhibit a spectrum distinct from thermal white noise, confirming the presence of the mCP.  However, to provide a comprehensive answer to the question, measurements on the output light need to be performed.
%\paragraph*{Spectra of the output field:}

The fluctuation spectrum of the light emitted from the cavity can be determined by using the two-frequency auto-correlation function
$\langle \delta \hat{J}^{\rm{out}}(\omega)  \delta\hat{J}^{\rm{out}}(\omega')\rangle =S^{\rm{out}}_{\rm{JJ}}(\omega, \omega') \delta(\omega+\omega')$ for $J=X, Y$. Then, using the input-output relation $\delta\hat J^{\rm{out}} = \sqrt{\kappa} \delta\hat J -\hat J_{\rm{in}}$ \cite{collett} and Eq.(\ref{vett}), the symmetric spectral function of the correlations of the output amplitude (phase) quadrature fluctuations $S^{\rm{out}}_{\rm{JJ}}(\omega)=(1/2\pi) \int d\omega' e^{-i(\omega-\omega')} \langle \delta \hat{J}^{\rm{out}}(\omega)  \hat{J}^{\rm{out}}(\omega')\rangle$ of the laser field in Fourier space is given by 
\begin{eqnarray}  \label{Sxx}
S^{\rm{out}}_{\rm{JJ}}(\omega)&=&\frac{1}{2}+\kappa \left\{\Gamma  |a_{\rm{J}}(\omega)|^2-\Re[b_{\rm{J}}(\omega) d(-\omega)]\right\} /  |d(\omega)|^2
\nonumber \\ 
&+&\frac{\kappa^2}{2} \left\{|b_{\rm{J}}(\omega)|^2 + |c_{\rm{J}}(\omega)|^2\right\} /  |d(\omega)|^2, 
\end{eqnarray}
where the exact expressions of $d(\omega)$, $a_{\rm{J}}(\omega)$, $b_{\rm{J}}(\omega)$, and $c_{\rm{J}}(\omega)$ for $J=X,Y$ are given in \cite{supmat}.
\begin{figure} [!]
\includegraphics[width=0.98 \columnwidth]{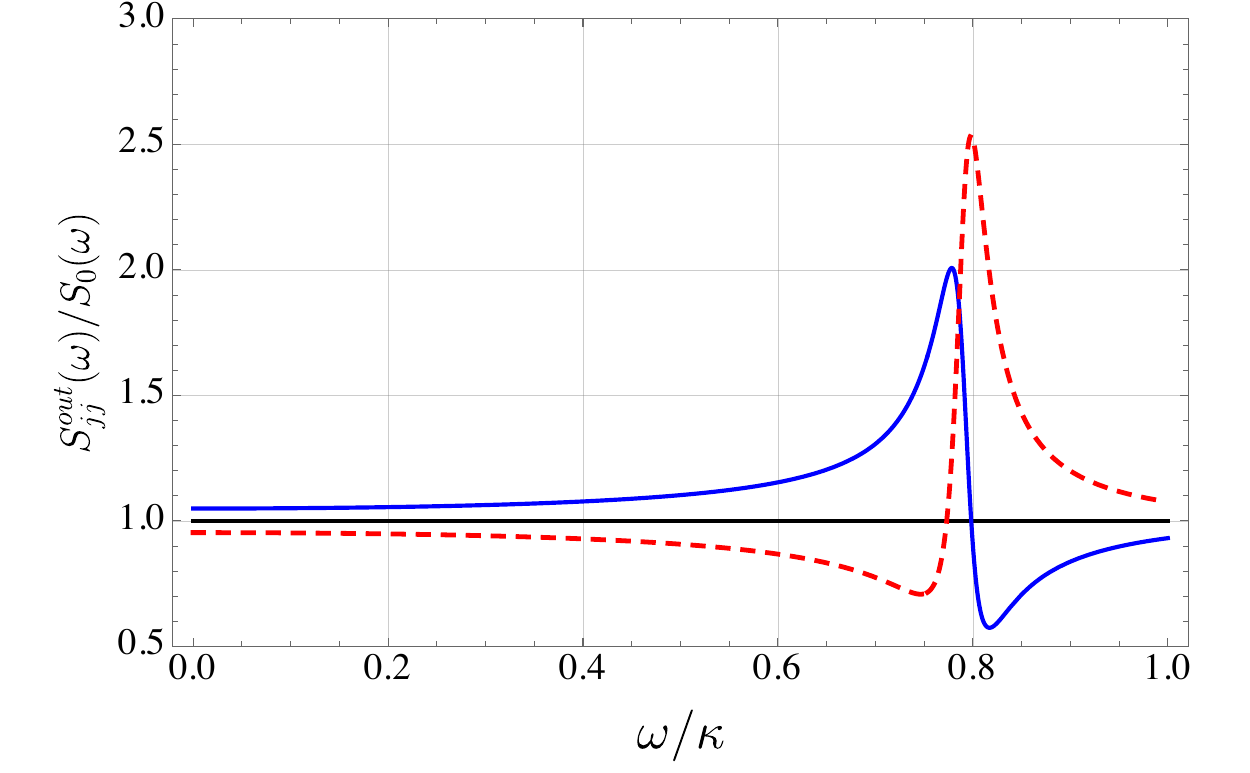}
\caption{(color online) The symmetric spectral function of the output field amplitude quadrature $S^{\rm{out}}_{\rm{XX}}(\omega)$ (blue curve) and phase quadrature $S^{\rm{out}}_{\rm{YY}}(\omega)$ (dashed red curve) fluctuations normalized to the spectrum of thermal noise $S^{\rm{out}}_0(\omega)$ (black line) are plotted as a function of $\omega/\kappa$ for a mCP charge of $q = 10^{-5}e_0$, a static ring field of ${\cal E}{\rm{x}} \simeq 7.25 \times 10^{10}$ V/m, a cavity field detuning of $\Delta_{0} = 0.8 \kappa$, a cavity length of $\mathcal{L} = 1$ cm, an input power of the trapping laser at $\mathcal{P} = 1.0$ mW, and a cavity finesse of $\mathcal{F} = 50000$.}
 %as a function of $\omega/\kappa$, for $q\,=\,10^{-5}e_0$ and a ring static field ${\cal E}_{\rm{x}} \, \simeq\, 7.25 \times 10^{10}$ V/m when cavity field detuning $\Delta_{0}\, =\, 0.8 \kappa$, cavity length $\mathcal{L} = 1$ cm, input power of the trapping laser $\mathcal{P} = 1.0$ mW and cavity finesse $\mathcal{F} = 50000$.}
\label{fig1}
\end{figure}
The effect of the charged ring is to modify the oscillation frequency $\omega_{\rm{m}}$, which gives rise to the effective $\Omega_{\rm{m}}$. Furthermore, due to the shifted position of the charged ring plane ($C_0 \neq 0$), when observing the output light, it no longer shows the thermal noise. An example of such a spectral function at temperature $T=300$ K and $P_{\rm{gas}}=10^{-10}$ Torr for $q=10^{-5}e_0$, $R=5.0$ mm, and a strong electrostatic field on the ring $\mathcal{E}_{\rm{x}} = 7.24986 \times 10^{10}$ V/m is reported in Fig. \ref{fig1} for $\Delta_{0} = 0.8 \kappa$ and $C_0 = \lambda$ (other examples are in \cite{supmat}). Hence, the measurement of the output light's spectrum \cite{supmat}, different from that of the thermal noise as shown in Fig. \ref{fig1}, will be the check for the existence of mCP in the NS considered.

\emph{What is now important is to provide a complete answer to the question posed in the title.}

%\emph{What is now of interest is to provide a complete answer to the question posed in the title.}
It is already evident from the figures that something non-classical is occurring. In fact, the amplitude and phase quadrature fluctuations of the output light exhibit squeezing at specific values of the output frequency. This finding could potentially signify the presence of a more significant quantum effect: the entanglement between the Gaussian states of light and the center of mass (CoM) position. To quantify the entanglement, here it is opted to use the logarithmic negativity \cite{VidalWerner,Plenio05, adesso}:
\begin{equation}
 E_n=\text{max}\{0, -\log(2\eta_-)\},
\end{equation}
where $\eta_-$ is the lowest symplectic eigenvalue of the partially transpose correlation matrix \cite{simon}.  
Indeed, the steady state of the bipartite quantum system formed by the mechanical mode of interest and the cavity mode can be fully characterized by their
correlation matrix. In fact, the quantum noises $\hat \eta(t)$ and ${\hat a}_{\rm{in}}(t)$ are quantum Gaussian noises with zero mean and the dynamics is linearized, consequently the stationary state of the system is a zero-mean bipartite Gaussian state, completely characterized by its $4\times 4$ correlation matrix $\bf V$ with elements
$V_{\rm{ij}}=[\langle u_{\rm{i}}(\infty)u_{\rm{j}}(\infty)+u_{\rm{j}}(\infty)u_{\rm{i}}(\infty)\rangle]/2$, which, when the stability conditions for the Eqs.(\ref{vett}) are satisfied \cite{supmat,stb}, they can be obtained  \cite{supmat} solving the Lyapunov equation \cite{lyapu,vitali3,vitalietal}
\begin{equation}
{\bf A \bf V} +{\bf V \bf {A}}^T = - {\bf D},
\end{equation}
where $\mathbf{A}$ is the drift matrix given in Eq.(\ref{matri}), and $D_{\rm{ij}} = \frac{1}{2}[\langle n_{\rm{i}}(\infty)n_{\rm{j}}(\infty)\rangle+\langle n_{\rm{j}}(\infty)n_{\rm{i}}(\infty)\rangle]\delta_{\rm{ij}}$ are the elements of the diagonal matrix $\mathbf{D}$ representing the stationary noises' correlation functions.
\begin{figure}[tb]
\includegraphics[scale=.41]{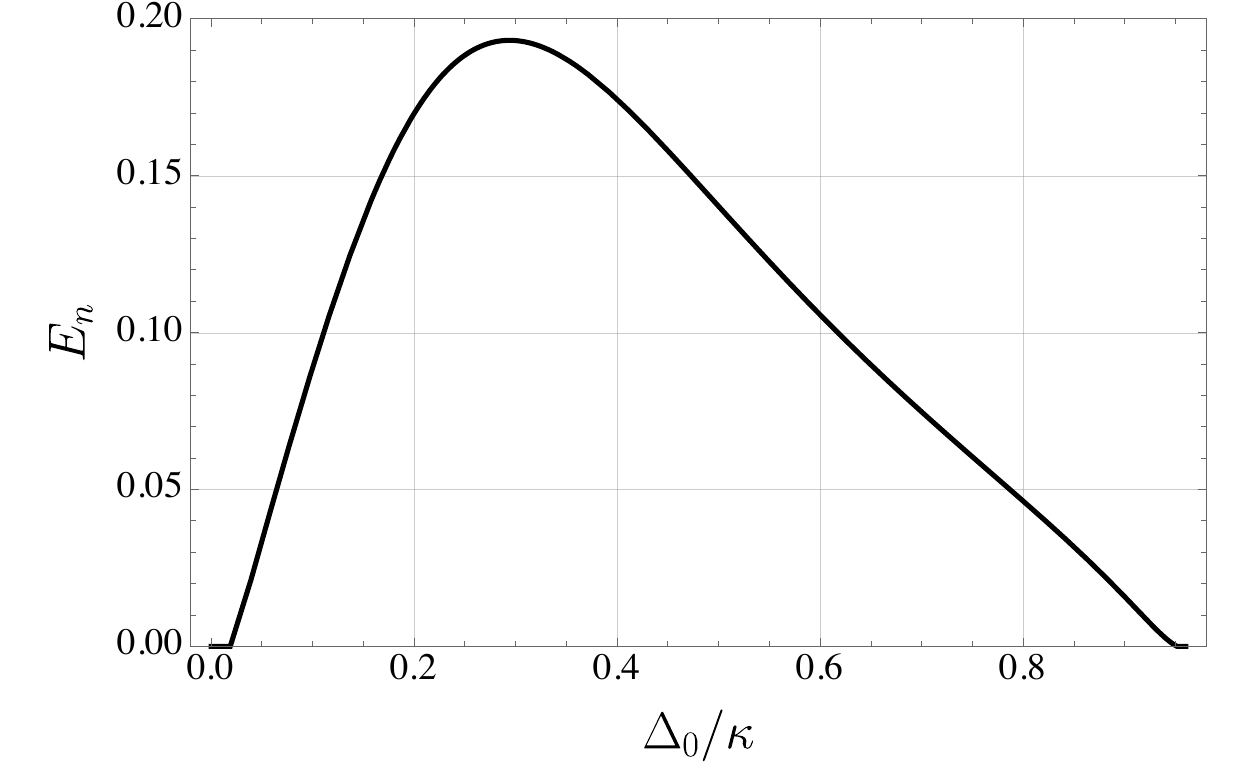}
\caption{The plot of the logarithmic negativity $E_n$ as a function of $\Delta_{0}/\kappa$ is shown for fixed values of $C_0= \lambda$ and $q= 10^{-5} e_0$. All other parameters are the same as those in Fig. \ref{fig1}.
%The plot of the logarithmic negativity $E_n$ as function of $\Delta_{0}/\kappa$ is shown considering fixed the values of $C_0= \lambda$, and $q= 10^{-5} e_0$. All other parameters are same as in Fig.\ref{fig1}.
}
\label{fig2}
\end{figure}

In clamped cantilever and similar cases, where the mechanical oscillation frequency is independent of the detuning, entanglement is usually observed by fixing the effective detuning at the Stokes sideband. In the scenario here considered, for any value of the mCP charge $q$, $\omega_{\rm{m}}$ depends on the detuning $\Delta_{0}$ and on $x_{\rm{s}}$. Moreover, the position of the droplet's CoM $x_{\rm{s}}$ depends on the detuning, the ring charge $Q$, and the shift constant $C_0$. Hence, one has to choose the best strategy to achieve better results for the entanglement. Considering all of that one obtains the best choice for the mCP charge $q=10^{-5}e_0$ ($Q\simeq\, 3.25$ C ) and an electrostatic field $\mathcal{E}_{\rm{x}} = 2.5 \times 10^{11}$ V/m gives a maximum $ E_n\simeq 0.2$ at $\Delta_{0}\simeq 0.3 \kappa$ as shown in Fig. \ref{fig2} \cite{supmat}. The results obtained here are for the entanglement inside the cavity. However, as it was already shown \cite{vitali3, vitali33}, by appropriately filtering the output light, the stationary entanglement between the internal mechanical mode and the output optical mode is higher at a selected output frequency. Recently, this ``long-standing prediction" was considered in Ref. \cite{aspelhammer}, where it is shown that in an appropriately pulsed regime, it is experimentally obtainable beyond the resolved sideband regime.

The strategy chosen is important in obtaining the entanglement, but one is limited to consider $\epsilon > 10^{-5}-10^{-6}$ because of the need for a very strong electrostatic field even at these values of the NS charge. Nevertheless, one could choose, for example, $\epsilon = 10^{-8}-10^{-11}$ and $\Delta(x_{\rm{s}}) \neq \omega_{\rm{m}}$. In this case, one obtains $x_{\rm{s}}$ from the transcendentalal Eq. (\ref{xsvalue}), and there is no more entanglement. However, one can still obtain an extremely feeble squeezing for the same value of the ring charge $Q$ as in the case of $\epsilon = 10^{-5}$, as shown in \cite{supmat}.

To obtain these results, assuming that every other possible noise or loss can be perfectly controlled and eliminated, the electrostatic field ${\cal E}_{\rm{x}}(x_{\rm{s}})$ should be as high as possible, but field emission from a rounded tip usually occurs at about $10^{10}$V/m in electron microscopy \cite{fursey}. Therefore, the smoothness of the metal ring must be considered, avoiding protruding parts and sharp points. It is also necessary to work at very low pressure on the order of $10^{-10} - 10^{-8}$ Torr while keeping the system stable, and this may be difficult to achieve at room temperature as considered here. However, this could be achieved at liquid Helium temperature where the Markov condition still holds, and similar results are obtained. Therefore, an electrostatic field value of this magnitude, or perhaps one or two orders of magnitude greater, could be adequately taken.

Despite all the possible difficulties, what has been shown is that, for values of the mCP not smaller than $q=10^{-6}e_0$ the complete answer to the question posed in the title is \emph{Yes}, if bound to matter mCPs exist, one can obtain quantum effects, i.e. squeezing on the output light and feeble entanglement between the internal light and the mechanical CoM mode, which are only due to the existence of the charged dark matter particle. For smaller mCP charge $q=\epsilon e_0$, at least down to $\epsilon \simeq 10^{-11}$ for $ \Delta(x_{\rm{s}}) \not= \omega_{\rm{m}}$ there is no more entanglement and the squeezing is vanishingly small, but the output light does not have a spectrum of the thermal noise. There will be a very narrow frequency peak \cite{supmat}. Whence one might expect to obtain a similar effect considering other possible very weak interactions of dark matter with our measurable world.

Acknowledgements.
P. Tombesi. would like to thank his wife Rita for her patience during Covid19 lockdown, when this work started, and Peter Zoller for his much appreciated clarifications. M. Asjad has been supported by the Khalifa University of Science and Technology under Award No. FSU-2023-014.


%\bibliography{QTDCe}
%%%%%%%%%%%%%%%%%%%%%%%%%%%%%%%%%%%%%%%%%%%%%%%%%%%%%%%%%%%%%%%%%%%%%%%%%%%%%%%%%%%%%%%%%%%%%%%%%%%%%%%%%%%%
%%%%%%%%%%%%%%%%%%%%%%%%%%%%%%%%%%%%%%%%%%%%%%%%%%%%%%%%%%%%%%%%%%%%%%%%%%%%%%%%%%%%%%%%%%%%%%%%%%%%%%%%%%%%
%%%%%%%%%%%%%%%%%%%%%%%%%%%%%%%%%%%%%%%%%%%%%%%%%%%%%%%%%%%%%%%%%%%%%%%%%%%%%%%%%%%%%%%%%%%%%%%%%%%%%%%%%%%%
%%%%%%%%%%%%%%%%%%%%%%%%%%%%%%%%%%%%%%%%%%%%%%%%%%%%%%%%%%%%%%%%%%%%%%%%%%%%%%%%%%%%%%%%%%%%%%%%%%%%%%%%%%%%
%%%%%%%%%%%%%%%%%%%%%%%%%%%%%%%%%%%%%%%%%%%%%%%%%%%%%%%%%%%%%%%%%%%%%%%%%%%%%%%%%%%%%%%%%%%%%%%%%%%%%%%%%%%%
%%%%%%%%%%%%%%%%%%%%%%%%%%%%%%%%%%%%%%%%%%%%%%%%%%%%%%%%%%%%%%%%%%%%%%%%%%%%%%%%%%%%%%%%%%%%%%%%%%%%%%%%%%%%
%%%%%%%%%%%%%%%%%%%%%%%%%%%%%%%%%%%%%%%%%%%%%%%%%%%%%%%%%%%%%%%%%%%%%%%%%%%%%%%%%%%%%%%%%%%%%%%%%%%%%%%%%%%%
\onecolumngrid

\newpage

\setcounter{equation}{0}
\setcounter{figure}{0}
\renewcommand{\theequation}{S.\arabic{equation}}
\renewcommand{\thefigure}{S.\arabic{figure}}
\makeatletter
\renewcommand\@biblabel[1]{[S.#1]}
\renewcommand\@cite[1]{[\colr{S}.#1]}
\makeatother

\hspace{-.5cm}
\begin{minipage}{18.2cm}

\begin{center}
{\bf\large  Supplemental material for 
``Does Bound to Matter Charged Dark Matter Exist in Bulk Materials and Can It Produce Observable Quantum Effects?''}

%\vspace{0.3cm}
\bigskip 
{\large Muhammad~Asjad$^1$, Paolo~Tombesi$^2$}

{\it $^1$Mathematics Department, Khalifa University of Science and Technology, 127788, Abu Dhabi, United Arab Emirates}\\
{\it  $^2$ School of Science and Technology, Physics Division, University of Camerino, 62032 Camerino (MC), Italy}

\smallskip
(Dated: \today)
\end{center}

\end{minipage}

\vspace{1cm}
%\twocolumngrid
\onecolumngrid

\section{system and equation of motions}
The Hamiltonian of the levitated silica droplet in one-sided optical cavity, in the simplified set-up considered in the main text, where only the trapping field is taken into account while the cooling field is not present with respect to \cite{S_kimble, S_isart, S_monteiro, S_aspelmeyer,S_kiesel},
 can be written as 
  \begin{eqnarray} \label{ham}
\hat{H_0}&= &\hbar\omega_{\rm{c}} \hat a^\dagger \hat a - \hbar g \cos^2(k x )\hat a^\dagger \hat a + \frac{ p_{\rm{x}}^2}{2 m} +\hbar E (\hat a e^{i\omega_{\rm{L}}t} + \hat a^\dagger e^{-i\omega_{\rm{L}}t}),
\end{eqnarray}
where $\omega_{\rm{c}}$ is the cavity angular frequency, $g$ the coupling constant between light and matter, $k=2\pi/\lambda$  the wave vector and  $E=\sqrt{\kappa \mathcal{P}/(\hbar \omega_{\rm{L}})}$  the amplitude of the driving/trapping electric field with $\kappa$ the cavity line width and $\mathcal{P}$ the power of  the input laser with angular frequency $\omega_{\rm{L}}$. The second term is the first order correction to the eigenvalue $\hbar \omega_{\rm{c}}$ of the energy inside the cavity. It is due to the small variation of the dielectric function of the cavity because of the existence of the dielectric nano-sphere (NS) of silica, considered as point-like in that context \cite{S_kimble, S_isart, S_monteiro}. It is obtained as \cite{S_joannopoulos}
\begin{equation}
\Delta \omega_{\rm{c}}=-\frac{\omega_{\rm{c}}\int d{\bf r}\Delta\varepsilon({\bf r})|{\bf E}({\bf r})|^2} {2 \int_{V_{\rm{c}}}d{\bf r}\varepsilon_0|{\bf E}({\bf r})|^2},
\end{equation}
where ${\bf E}({\bf r})=E \,\hat{\bf y}e^{-\frac{y^2+z^2}{w^2}}\cos(kx-\varphi)$ the TEM00 mode profile polarized along the unity vector $\hat{\bf y}$ normalized to the mode volume $V_{\rm{c}}=\pi w^2{\cal L}/4$ with $w$ the mode's waist and $\cal L$ the cavity length. As assumed in the main text, the spread in the transverse directions is considered much smaller then the mode waist, then $e^{-\frac{y^2+z^2}{w^2}}\simeq 1$. The angular frequency of the mode is $\omega_{\rm{c}}$  and  the induced dipole moment is 
${\bf p}=\Delta\varepsilon({\bf r}){\bf E}\simeq \alpha_{\mathrm{ind}}\delta({\bf r}-{\bf r}'){\bf E}$ for the point-like dipole with $\alpha_{\mathrm{ind}}=3\varepsilon_0V_{\rm{s}}\frac{\varepsilon-1}{\varepsilon+2}$ \cite{S_stratton, S_kimble} where $V_{\rm{s}}$ is the NS volume with radius $r$, and $\varepsilon_0$ the vacuum dielectric constant. Hence the coupling constant $g=\frac{3V_{\rm{s}}}{2V_{\rm{c}}}\frac{\varepsilon-1}{\varepsilon+2}\omega_{\rm{c}}$ and it has been chosen the phase $\varphi=0$.

The center of mass (CoM) position and momentum operators of the NS are denoted as $x$ and $p$, respectively. They satisfy the canonical commutation relation $(x,p)=i\hbar$. The operator $\hat a (\hat a^\dagger)$ represents the annihilation (creation) operator of the cavity mode, with commutation relation $(\hat a,\hat a^\dagger)=1$. The third term in the equation represents the kinetic energy of the NS, which has a mass of $m=\rho V_{\rm{s}}$, where $\rho$ is the mass density of the sphere. The NS is trapped at an antinode of the standing wave trapping field, exhibiting an oscillation frequency of $\omega_{\rm{m}}/2\pi$. Interestingly, contrary to conventional opto-mechanical systems, this oscillation frequency is primarily influenced by the detuning between the frequency of the driving laser and the cavity frequency, as demonstrated later in this discussion \cite{S_monteiro}.

The reference frame is established with its origin at the equilibrium position of the NS center of mass, aligning the x-axis with the cavity axis, which coincides with an antinode of the trapping field. Under the assumption of a charged dark matter particle existing within the NS, possessing a milli-charge $q=\epsilon e_0$ (with $\epsilon$ ranging from $10^{-2}$ to $10^{-16}$ \cite{S_Ahlers}, and $e_0$ representing the charge on an electron), a strategy is devised to amplify its interaction with our observable world. This strategy involves the strategic placement of a highly charged metallic ring.

This metallic ring, characterized by a radius $R\,>>\,r$, is positioned in a plane orthogonal to the x-axis. The center of the ring is situated at a distance $C_0$ from the origin. Due to the symmetry of the configuration, the electrostatic field generated by the ring, featuring an electric charge $Q=2\pi l R$ where $l$ is the linear charge density, possesses non-zero values solely along the cavity axis. Precisely, at the position $x$ of the point-like NS, the electrostatic field is significant. It can be described by the expression:
\begin{equation}
\mathcal{E}_{\rm{x}}(x)=\frac{Q(C_0+x)}{4\pi\epsilon_0 R^3\left[1+\left(\frac {C_0+x}{R}\right)^2\right]^{3/2}}.
\end{equation}

This field contributes to the total energy, yielding:
\begin{equation}
q\phi(x)= \frac{qQ}{4\pi \epsilon_0 R}\left[1+\left(\frac{C_0+x}{R}\right)^2 \right]^{-\frac{1}{2}},
\end{equation}
where $\phi(x)$ represents the electrostatic potential and $q$ corresponds to the millicharge. For the point charge NS, the associated Hamiltonian is given by:
\begin{equation}
H_{NS}=\frac{1}{2m}({\bf p}-q{\bf A})^2+q\phi(x) \simeq \frac{p^2_{\rm{x}}}{2m} + q \phi(x).
\end{equation}
In this expression, the motion is considered only along the \emph{x-axis}, and the cavity's electromagnetic field is polarized along the $y$ axis. As a result, the scalar product ${\bf A}\cdot {\bf p}=0$, and the quadratic term is negligible due to the smallness of $q$. Consequently, the primary contribution to Eq.(\ref{ham}) arises from the electrostatic potential. Transitioning to the rotating frame at the pump frequency $\omega_{\rm{L}}/2\pi$ using the rotation operator $\hat R= e^{i\hat a^\dagger\hat a \omega_{\rm{L}} t}$ yields:
\begin{equation}
\hat H =- i\hbar \hat R \dot{\hat R}^\dagger + \hat R(\hat H_0+q\phi(x))\hat R^\dagger,
\end{equation}
which gives rise to the Hamiltonian (discussed in the main text)
  \begin{eqnarray} \label{rotham}
  \hat H&=&-\hbar \Delta_{0} \hat a^\dagger\hat a - \hbar g \cos^2(k x )\hat a^\dagger \hat a 
   + \frac{ p^2}{2 m} + q\phi(x) +\hbar E (\hat a^\dagger + \hat a),
  \end{eqnarray}
where $\Delta_{0}=\omega_{\rm{L}} -\omega_{\rm{c}}$ the detuning (on $p$ the subscript $x$ was eliminated). The system's nonlinear Heisenberg-Langevin equations of motion are
 \begin{eqnarray} \label{eqsmot}
 \dot x&=& \frac{p}{m}, \\ \nonumber
 \dot p&=&\hbar g \hat a^\dagger\hat a \frac {d}{d x}\cos^2(kx) -q\frac {d}{d x}\phi(x)-\frac {\gamma}{2}p+\xi(t), \\ \nonumber
 \dot {\hat a}&= &\left[i\{\Delta_{0}+g\cos^2(kx)\}- \frac{\kappa}{2}\right] \hat a-i E +\sqrt{\kappa}\, \hat{a}_{\rm{in}}(t),
 \end{eqnarray}
where the damping constants $\kappa$ and $\gamma$ respectively for the cavity mode and the mechanical mode have been introduced. The Hermitian stochastic force $\xi(t)$ has expectation $\langle\xi(t)\rangle=0$ and correlations 
 \begin{equation} \label{corr}
 \frac{1}{2}[\langle\xi(t)\xi(t')\rangle+\langle\xi(t')\xi(t)\rangle]=\tilde{\Gamma} \delta(t-t')
 \end{equation}
with dimensions as the square of a force, and $\tilde{\Gamma}$ as obtained below. However, Eq.(\ref{corr}) is, in reality, valid for Markovian processes \cite{S_giovannettivitali} and this holds true as long as $\hbar \omega_{\rm{m}} \ll k_{\rm{B}}T$ or when $\omega_{\rm{m}}/\gamma \gg 1$ \cite{S_kac}. Consequently, one can treat the process as Markovian. Nevertheless, it's important to note that mechanical modes cease to exhibit Markovian behavior at very low temperatures, as indicated in \cite{S_haake}.
Now, introducing the displacements $\hat a=a_{\rm{s}} + \delta\hat a$ and $x=x_{\rm{s}}+\delta x, \,\,p=p_s+\delta p$, with $a_{\rm{s}}$, $x_{\rm{s}}$ and  $p_s$ the new steady state mean values, one can linearize the equations of motion up to the first order in fluctuations with respect to the steady state values, and the system of Eqs.(\ref{eqsmot}) becomes
\begin{eqnarray} \label{lineqs}
\delta\dot x&=&\frac{\delta p}{m},\nonumber \\ 
\delta\dot p&=&-(m \omega_{\rm{m}}^2+A_{\rm{q}})\delta x -\sqrt 2 \hbar g a_{\rm{s}} k\sin(2kx_{\rm{s}}) \delta \hat X-\frac {\gamma}{2}\delta p+\xi(t),\nonumber\\ 
\delta \dot{\hat a}&=&i\{\Delta (x_{\rm{s}})+i \frac{\kappa}{2}\} \delta \hat a - iga_{\rm{s}} k \sin(2kx_{\rm{s}}) \delta x + \sqrt \kappa\, \hat a_{\rm{in}},
\end{eqnarray}
where
\begin{equation}
A_{\rm{q}}=\frac{q Q }{4\pi\epsilon_0 R^3} \left[1-2\left\{\dfrac{C_0+x_{\rm{s}}}{R}\right\}^2\right] \left[1+\left\{\dfrac{C_0+x_{\rm{s}}}{R}\right\}^2\right]^{-5/2}\,\simeq \frac{q Q }{4\pi\epsilon_0 R^3} 
\end{equation}
and $\Delta(x_{\rm{s}})=\Delta_{0}+g\cos^2kx_{\rm{s}}$ is the effective detuning, which depends on the steady state position of NS CoM, and $\delta\hat X=(\delta \hat a + \delta \hat a^\dagger)/\sqrt 2$ the amplitude quadrature of the internal optical fluctuation, and we considered $(C_0+x_{\rm{s}})<<R$. Furthermore, we have introduced the mechanical angular frequency oscillation  
\begin{equation} \label{osc}
\omega_{\rm{m}}= \sqrt{2\hbar g k^2 |a_{\rm{s}}|^2 \cos(2kx_{\rm{s}})/m}.
\end{equation}
Since it is always possible to chose a reference phase such that $a_{\rm{s}}$ is real, the values of $a_{\rm{s}}$
is given by
\begin{equation}\label{alpha}
%a_{\rm{s}} =\sqrt{|a_{\rm{s}}|^2}=\sqrt{\frac{4|E|^2}{4(\Delta_{0}+g\cos^2kx_{\rm{s}})^2+\kappa^2}}
a_{\rm{s}} =\sqrt{\frac{4|E|^2}{4(\Delta_{0}+g\cos^2kx_{\rm{s}})^2+\kappa^2}}
\end{equation}
and $x_{\rm{s}}$ is given solving the following transcendental equation, obtained eliminating the terms non depending on fluctuations in getting Eqs.(\ref{lineqs}):
\begin{eqnarray} \label{S.steady}
A_{\rm{q}}(x_{\rm{s}}+C_0)+\frac{4\hbar g k |E|^2 \sin(2kx_{\rm{s}})}{4(\Delta_{0}+g\cos^2kx_{\rm{s}})^2+\kappa^2} &=&0,
\end{eqnarray} 
whose solution fix the values of $x_{\rm{s}}$  once choosing  $\Delta_{0}$, the values of the ring's parameters, and the mCP charge.

In opto-mechanical systems with clamped cantilever or membrane in the middle and similar, usually the detuning is fixed by the Pound-Drever-Hall technique, to compensate the spring constant shift or to fix the effective detuning at a chosen value. However here, the effective detuning depends on the new CoM position $x_{\rm{s}}$, which depends not only on $C_0$ but, as the average photons number within the cavity $|a_{\rm{s}}|^2$, and moreover also the angular frequency $\omega_{\rm{m}}$, all  of them depend on the detuning $\Delta_{0}$. 
Thus, upon setting the value of $\Delta_{0}$, all relevant quantities are obtained by using their corresponding equations (Eqs. \ref{osc}, \ref{alpha}). Although the effective detuning can be chosen by the experimenter, we maintain the consideration of it as not fixed in this context. This choice of $\Delta_{0}$ holds significance in determining other parameters. For the stability of the mechanical motion (second equation of Eqs.(\ref{lineqs})), which should be satisfied also in absence of any mCP, the condition $\cos(2kx_{\rm{s}}) \ge 0$ should be satisfied, furthermore it should be $C_0\not=0$ otherwise the mechanical and the optical motions are totally decoupled so no information could be obtained with this setup from the output field. All of that imposes constraints: i.e. the steady state value of the CoM position of NS should satisfy $-\frac{\pi}{4k}<x_{\rm{s}}<\frac{\pi}{4k}$ with $x_{\rm{s}}\not=0$ for  $C_0\not=0$  when solving Eq. (\ref{S.steady}).
To have all fluctuation operators dimensionless it is worth introducing the canonical transformation 
 %\begin{equation}
$ \delta x = \sqrt\frac{\hbar}{m\omega_{\rm{m}}} \delta\hat x$
and $ \delta p= \sqrt{\hbar m \omega_{\rm{m}}} \delta\hat p$
 %\end{equation}
 with commutation relation $( \delta\hat x, \delta\hat p)=i$. Finally, introducing also the internal optical phase quadrature fluctuation $\delta\hat Y=(\delta \hat a - \delta \hat a^\dagger)/i\sqrt 2$ and defining $G=\sqrt{2}
  g k a_{\rm{s}}\sin(2kx_{\rm{s}})\sqrt{\frac{\hbar}{m\omega_{\rm{m}}}}$ Eq.(\ref{lineqs}) becomes
 \begin{equation} \label{linvet}
 \bf{\dot{\hat u}}(t)=\bf{A} \bf{\hat u}(t)+\bf{\hat n}(t),
 \end{equation}
where $\bf A$ is the drift matrix, which is given by
 \begin{equation}\label{matris}
{\bf A} = \begin{pmatrix}
0 &\omega_{\rm{m}} & 0 &0 \\
-\Omega_{\rm{m}} & -\frac{\gamma}{2}& -G & 0\\
0 & 0& -\frac{\kappa}{2} & \Delta (x_{\rm{s}})
\\-G & 0 & -\Delta (x_{\rm{s}}) & -\frac{\kappa}{2}
\end{pmatrix}.
\end{equation}
with $\Omega_{\rm{m}} = \omega_{\rm{m}} + \frac{A_{\rm{q}}}{m\omega_{\rm{m}}}$ represents the effective angular frequency. The vector ${\bf{\hat u}}(t) = (\delta\hat x, \delta\hat p, \delta\hat X, \delta\hat Y)^T$ denotes the quadratures vector, and the corresponding noise vector is ${\bf{\hat n}}(t) = (0, \hat\eta(t), \sqrt{\kappa}\hat X_{\rm{in}}(t), \sqrt{\kappa}\hat Y_{\rm{in}}(t))^T$ (the superscript T indicates transposition). Here, $\hat X_{\rm{in}}$ and $\hat Y_{\rm{in}}$ respectively denote the amplitude and phase quadratures of shot noise. The new stochastic term is $\hat\eta(t)=\hat \xi(t)/\sqrt{\hbar m \omega_{\rm{m}}}$  with expectation $\langle\hat\eta(t)\rangle=0$ and correlation $\langle\hat\eta(t)\hat\eta(t')\rangle=\Gamma\delta(t-t')$  with $\Gamma=\tilde{\Gamma}/(\hbar m \omega_{\rm{m}})$, and one has $\tilde\Gamma=\hbar m \omega_{\rm{m}} \gamma(n_0+{1/ 2})$, where $n_0=(e^{\hbar\omega_{\rm{m}}/k_{\rm{B}}T}-1)^{-1} \approx k_{\rm{B}}T/\hbar\omega_{\rm{m}}$ hence $\tilde\Gamma \simeq \gamma m k_{\rm{B}}T$. The optical shot noise is instead well described as a Markovian process and considering  the annihilation (creation) operator $\hat a_{\rm{in}}(t)$($\hat a^\dagger_{\rm{in}}(t))$ it has expectation $\langle \hat a_{\rm{in}}(t)\rangle=\langle \hat a^\dagger_{\rm{in}}(t)\rangle=0$, and correlations $\frac{1}{2}[\langle\hat a_{\rm{in}}(t)\hat a_{\rm{in}}^\dagger(t')\rangle+\langle\hat a_{\rm{in}}(t')\hat a_{\rm{in}}^\dagger(t)\rangle]= (n_{\rm{in}}+1)\delta(t-t')$ and $\frac{1}{2}[\langle\hat a_{\rm{in}}^\dagger(t)\hat a_{in}(t')\rangle+\langle\hat a_{\rm{in}}^\dagger(t')\hat a_{\rm{in}}(t)\rangle]= n_{\rm{in}}\delta(t-t')$, with $n_{\rm{in}}=(e^{\hbar \omega_{\rm{L}}/k_{\rm{B}} T}-1)^{-1}$ the average number of input noise photons at the laser frequency, which could be considered zero as long as $\hbar \omega_{\rm{L}} >> k_{\rm{B}} T$ and this always happens at optical frequency (being $k_{\rm{B}}$ the Boltzmann constant and $T$ the equilibrium bath's temperature).
 
To enhance rigor, the value of $\tilde\Gamma$ is derived by invoking the fluctuation-dissipation theorem (F-D) \cite{S_kubo}, which remains applicable even in the presence of a potential term. At thermal equilibrium, this theorem establishes a link between damping and diffusion. Considering the second equation of Eq.(\ref{eqsmot}), we find that for a fixed number of cavity photons, denoting $\tilde\Gamma$ as the diffusion constant, the steady-state conditional probability distribution of momentum can be obtained by employing the following Fokker-Planck equation for the conditional probability distribution of momentum \cite{S_gardiner}:

\begin{equation}
\frac{\partial}{\partial t}W(p,t;p_0,t_0)=\frac{\partial}{\partial p}({\dfrac{\gamma}{2}}p+ {\dfrac{\tilde\Gamma}{2}}\frac{\partial}{\partial p})W(p,t;p_0,t_0).
\end{equation}
The steady-state solution is then given by $W(p,\infty;p_0,t_0) \propto e^{-{\gamma p^2\tilde \Gamma}/2}$. In accordance with the F-D theorem at thermal equilibrium, this solution should align with the Maxwell-Boltzmann distribution $\propto e^{-\frac{\langle p^2\rangle}{2m k_{\rm{B}} T}}$, leading to the relationship $\tilde\Gamma=\gamma m k_{\rm{B}} T$. For the mechanical mode, two diffusion terms are present (i.e. $\tilde \Gamma= \tilde\Gamma_{\rm{ph}}+ \tilde\Gamma_{\rm{gas}}$). $\tilde\Gamma_{\rm{ph}}$ emerges from the scattering of trapping field photons \cite{S_kimble}, and the other, $\tilde\Gamma_{\rm{gas}}$, arises from the scattering of residual gas molecules. Consequently, the F-D theorem at thermal equilibrium gives rise to two distinct dissipative channels, $\gamma=\gamma_{\rm{ph}}+\gamma_{\rm{gas}}$, attributed respectively to the scattering of photons striking the droplet and the interactions involving residual gas molecules. In \cite{S_kimble}, a method is presented for obtaining the diffusive term as follows:
$
\Gamma_{\rm{ph}}=\frac{4\pi^2}{5} \frac{\varepsilon-1}{\varepsilon+2}\left[\frac{V_{\rm{s}}}{\lambda^3}\right]\omega_{\rm{m}}
$
Hence, in this context, $ \tilde\Gamma_{\rm{ph}}=\hbar m \omega_{\rm{m}}\Gamma_{\rm{ph}}=\gamma_{\rm{ph}} m k_{\rm{B}} T$ \cite{S_kimble}, leading to $\gamma_{\rm{ph}}= \Gamma_{\rm{ph}}\frac{\hbar \omega_{\rm{m}}}{ k_{\rm{B}} T}$. Regarding the other dissipative channel, the kinetic theory of gas, as elaborated in \cite{S_isart}, yields $\gamma_{\rm{gas}}=\frac{4\pi r^2 P_{\rm{gas}}}{m v}$, where $v=\sqrt{3 k_{\rm{B}} T/m_{\rm{a}}}$ denotes the mean velocity of gas molecules and $m_{\rm{a}}$ represents the mass of a gas molecule. Consequently, the diffusion constant $\tilde\Gamma_{\rm{gas}}=\gamma_{\rm{gas}} m k_{\rm{B}} T$, or equivalently, $\Gamma_{\rm{gas}}=\gamma_{\rm{gas}} \frac{k_{\rm{B}} T}{\hbar \omega_{\rm{m}}}$. Then the total damping and diffusion constants of mechanical mode are given by
\begin{eqnarray}
\gamma&= &\frac{4\pi^2}{5} \frac{\varepsilon-1}{\varepsilon+2}\left[\frac{V_{\rm{s}}}{\lambda^3}\right]\omega_{\rm{m}}\frac{\hbar \omega_{\rm{m}}}{ k_{\rm{B}} T}+\frac{4\pi r^2 P_{\rm{gas}}}{m v}, \\ \nonumber
\Gamma&=&\gamma \frac{k_{\rm{B}} T}{\hbar \omega_{\rm{m}}}.
\end{eqnarray}
It is evident that the photonics contribution to damping is nearly always negligible, only becoming significant at extremely low temperatures (below mK), a range where the validity of the position in Eq.(\ref{corr}) is compromised \cite{S_giovannettivitali}. On the other hand, damping due to air viscosity is pressure-dependent and scales with temperature as $T^{-1/2}$. Concerning diffusion, the contribution from photon scattering remains constant, dependent solely on the mechanical frequency, while the contribution from air viscosity is pressure-dependent and scales with temperature as $T^{1/2}$. 

The prerequisite for the effectiveness of our scheme is the stability of the system. System stability is realized when all eigenvalues of the drift matrix ${\bf A}$ possess negative real parts, i.e., ${\bf A}-\lambda {\bf I} = 0$. According to the the Routh-Hurwitz criteria \cite{S_stb}, the system maintains stability ($\Re [\lambda] < 0$) and reaches its steady state only if the following two stability conditions are simultaneously satisfied:
\begin{eqnarray}
S_1&=& \omega_{\rm{m}} \Omega_{\rm{m}}(4\Delta^2(x_{\rm{s}})+\kappa^2)-4 G^2 \Delta(x_{\rm{s}}) \omega_{\rm{m}}>0,\\
S_2&=& 2(\gamma+2\kappa) \left[ \left\{\kappa  (4\Delta^2(x_{\rm{s}})+(\gamma+ \kappa)^2 ) +2\gamma\omega_{\rm{m}}\Omega_{\rm{m}} \right\} 
\left\{ \gamma (4 \Delta^2(x_{\rm{s}}) +\kappa^2)+8 \kappa \omega_{\rm{m}} \Omega_{\rm{m}} \right\}  \right. \nonumber\\
&-&\left.
 2 (\gamma+2\kappa)^2\{\omega_{\rm{m}} \Omega_{\rm{m}}(4\Delta^2(x_{\rm{s}})+\kappa^2)-4 G^2 \Delta(x_{\rm{s}}) \omega_{\rm{m}} \}  \right]>0.
\end{eqnarray}
In the following sections, we will chose parameters that ensure the two stability conditions mentioned above are met.
\section{spectra of out-put field}
To address the question posed in the title of the main text, a measurement must be carried out on the sole possible mode that can exit the cavity, namely, the optical mode. To facilitate this, let's introduce the Fourier Transforms $f(t)=\int_{-\infty}^\infty \frac{d\omega}{\sqrt{2\pi}}e^{-i\omega t}f(\omega)$. With this, the system described by Eq.(\ref{linvet}) can be expressed as:
\begin{equation}\label{vetts}
-(i\omega {\bf I}+ {\bf A}) {\bf \hat u}(\omega)= {\bf \hat n}(\omega),
\end{equation}
where ${\bf I}$ denotes the $4 \times 4$ identity matrix, $\bf A$ is as defined in Eq.(\ref{matris}), and ${\bf \hat u}(\omega)=({\delta\hat x}(\omega),{\delta\hat p}(\omega),{\delta\hat X}(\omega),{\delta\hat Y}(\omega))^T$, while ${\bf \hat n}(\omega)=(0,\hat\eta(\omega),\sqrt\kappa \hat X_{\rm{in}}(\omega),\sqrt\kappa \hat Y_{\rm{in}}(\omega))^T$. In the context mentioned in the main text, where $C_0=0$, the matrix $\bf A$ simplifies to:
\begin{equation}
{\bf A_0}=\begin{pmatrix}0&\omega_{\rm{m}}&0&0\\-\Omega_{\rm{m}}&-\frac{\gamma}{2}&0&0\\0&0&-\frac{\kappa}{2}&\Delta_{0}\\0&0&-\Delta_{0}&-\frac{\kappa}{2}
\end{pmatrix}.
\end{equation}
In this scenario, the solution for phase and amplitude quadrature fluctuations of the internal optical mode becomes independent of $\delta\hat x_0(\omega)$:
\begin{equation}
\delta \hat Y_0(\omega)=\sqrt\kappa\frac{-\Delta_{0}\hat X_{\rm{in}}(\omega)+(\frac{\kappa}{2}-i \omega)\hat Y_{\rm{in}}(\omega)}{\Delta_{0}^2+(\frac{\kappa}{2}-i \omega)^2},
\end{equation}
while for $\delta\hat x_0$:
\begin{equation}
\delta\hat x_0=\frac{\omega_{\rm{m}}}{\omega_{\rm{m}} \Omega_{\rm{m}}-\omega^2-i\omega \frac{\gamma}{2}} \hat\eta(\omega).
\end{equation}
Consequently, the quadrature fluctuations of the output optical field, defined as $\delta \hat Y_0^{\rm{out}}(\omega)=\sqrt\kappa \delta \hat Y_0(\omega)-\hat Y_{\rm{in}}(\omega)$, become independent of $\delta \hat x_0(\omega)$. As a result, no information can be extracted without introducing another optical mode, as demonstrated in the references cited in the main text. The expression for $\delta \hat Y_0^{\rm{out}}(\omega)$ is as follows:
\begin{eqnarray}
\delta \hat Y_0^{\rm{out}}(\omega)&=&\frac{-\kappa\Delta_{0}}{\Delta_{0}^2+(\frac{\kappa}{2}-i \omega)^2}\hat X_{\rm{in}}(\omega)
+\frac{\omega^2+\frac{\kappa^2}{4}-\Delta_{0}^2}{\Delta_{0}^2+(\frac{\kappa}{2}-i \omega)^2}\hat Y_{\rm{in}}(\omega).
\end{eqnarray}
It is evident that the symmetric spectrum of phase quadrature fluctuation, defined as $S_{Y_0Y_0}^{\rm{out}}(\omega,\omega')=\frac{1}{2}[\langle\delta \hat Y_0^{\rm{out}}(\omega)\delta \hat Y_0^{\rm{out}}(\omega')\rangle+\langle\delta \hat Y_0^{\rm{out}}(\omega')\delta \hat Y_0^{\rm{out}}(\omega)\rangle],$ results in:
\begin{equation}
S_{Y_0Y_0}^{\rm{out}}(\omega,\omega')=\frac{1}{2}\delta(\omega+\omega').
\end{equation}
A similar outcome is observed for the amplitude quadrature fluctuation. Instead, going back to the full Eq.(\ref{vetts}) after some algebra one obtains
\begin{eqnarray}
\delta\hat X(\omega)&=&-G \Delta(x_{\rm{s}})\chi_{\rm{c}}(\omega)\delta\hat x(\omega)  +\Delta(x_{\rm{s}})\chi_{\rm{c}}(\omega)\sqrt{\kappa}\hat X_{\rm{in}}(\omega) +(\frac{\kappa}{2}-i \omega)\chi_{\rm{c}}(\omega)\sqrt{\kappa}\hat Y_{\rm{in}}(\omega),\\
%%%%%%%%%%%%%%%%
\delta\hat Y(\omega)&=&-G \Delta(x_{\rm{s}})\chi_{\rm{c}}(\omega)\delta\hat y(\omega) +\Delta(x_{\rm{s}})\chi_{\rm{c}}(\omega)\sqrt{\kappa}\hat Y_{\rm{in}}(\omega) -(\frac{\kappa}{2}-i \omega)\chi_{\rm{c}}(\omega)\sqrt{\kappa}\hat X_{\rm{in}}(\omega),
\end{eqnarray}
where $\chi_{\rm{c}}^{-1}(\omega)=(\Delta^2(x_{\rm{s}})+(\frac{\kappa}{2}-i \omega)^2)$. It turns out, however, that it is better to write 
\begin{eqnarray}
\delta \hat X(\omega)= \frac{\hat\eta(\omega)a_{\rm{X}}(\omega)+\sqrt{\kappa} \hat X_{\rm{in}}(\omega)b_{\rm{X}}(\omega)+\sqrt{\kappa} \hat Y_{\rm{in}}(\omega)c_{\rm{X}}(\omega)}{d(\omega)},\\
%%%%%%%%%%%%%%%%%%%%%%%%%
\delta \hat Y(\omega)= \frac{\hat\eta(\omega)a_{\rm{Y}}(\omega)+\sqrt{\kappa} \hat X_{\rm{in}}(\omega)b_{\rm{Y}}(\omega)+\sqrt{\kappa} \hat Y_{\rm{in}}(\omega)c_{\rm{Y}}(\omega)}{d(\omega)},
\end{eqnarray}
where
\begin{eqnarray}
a_{\rm{X}}(\omega)&=&G \omega_{\rm{m}} \Delta(x_{\rm{s}}) \\ 
b_{\rm{X}}(\omega)&=&(\frac{\kappa}{2}-i \omega)\chi_{\rm{m}}^{-1}(\omega)\\
c_{\rm{X}}(\omega)&=&\Delta(x_{\rm{s}})\chi_{\rm{m}}^{-1}(\omega)\\
a_{\rm{Y}}(\omega)&=&G \omega_{\rm{m}} (\frac{\kappa}{2}-i \omega) \\ 
b_{\rm{Y}}(\omega)&=&-\Delta(x_{\rm{s}})\chi_{\rm{m}}^{-1}(\omega)+\omega_{\rm{m}} G^2 \\ 
c_{\rm{Y}}(\omega)&=&(\frac{\kappa}{2}-i \omega)\chi_{\rm{m}}^{-1}(\omega),
\end{eqnarray}
where $\chi_{\rm{m}}(\omega)=( \omega_{\rm{m}}\Omega_{\rm{m}} - \omega^2  - i\omega \frac{\gamma}{2})^{-1},$ and
\begin{equation}
{d}(\omega)=[\chi_{\rm{c}}^{-1}(\omega)\chi_{\rm{m}}^{-1}(\omega)-G^2\omega_{\rm{m}}\Delta(x_{\rm{s}})]
\end{equation} 
is the determinant of $-(i\omega {\bf I}+ {\bf A})$. Now both $\delta \hat X(\omega)$ and  $\delta \hat Y(\omega)$ depend on $\delta \hat x(\omega)$ in some cumbersome way. Taking into account the output quadrature fluctuations $\delta \hat J^{\rm{out}}(\omega)=\sqrt\kappa\delta \hat J(\omega)-\hat J_{\rm{in}}(\omega)$ for $(J=Y,X)$, and $\langle \delta \hat J^{\rm{out}}(\omega)\rangle=0$, the symmetric spectra of output quadrature fluctuations $S^{\rm{out}}_{\rm{JJ}}(\omega, \omega')=\frac{1}{2}\langle \delta\hat J^{\rm{out}}(\omega)\delta\hat J^{\rm{out}}(\omega')+\delta\hat J^{\rm{out}}(\omega')\delta\hat J^{\rm{out}}(\omega)\rangle$ after some algebra are 
\begin{eqnarray} \label{syy}
S^{\rm{out}}_{YY}(\omega,\omega')&=&\left\{\frac{1}{2}+\kappa\Gamma \frac{|a_{\rm{Y}}(\omega)|^2}{|d(\omega)|^2}+\frac{1}{2}\left[\kappa^2\big(\frac{|b_{\rm{Y}}(\omega)|^2}{|d(\omega)|^2}+\frac{|c_{\rm{Y}}(\omega)|^2}{|d(\omega)|^2}\big) 
-\kappa\frac{\Re[c_{\rm{Y}}(\omega) d(-\omega)]}{|d(\omega)|^2}\right]\right\} \delta(\omega+\omega'),
\end{eqnarray}
\begin{eqnarray} \label{sxx}
S^{\rm{out}}_{\rm{XX}}(\omega,\omega')&=&\left\{\frac{1}{2}+\kappa\Gamma \frac{|a_{\rm{X}}(\omega)|^2}{|d(\omega)|^2}+\frac{1}{2}\left[\kappa^2\big(\frac{|b_{\rm{X}}(\omega)|^2}{|d(\omega)|^2}+\frac{|c_{\rm{X}}(\omega)|^2}{|d(\omega)|^2}\big) 
-\kappa\frac{\Re[b_{\rm{X}}(\omega) d(-\omega)]}{|d(\omega)|^2}\right]\right\} \delta(\omega+\omega'),
\end{eqnarray}
where $d(\omega)^*=d(-\omega)$ ($\Re[...]$ is the real part).
\begin{figure}
\includegraphics[scale=.58]{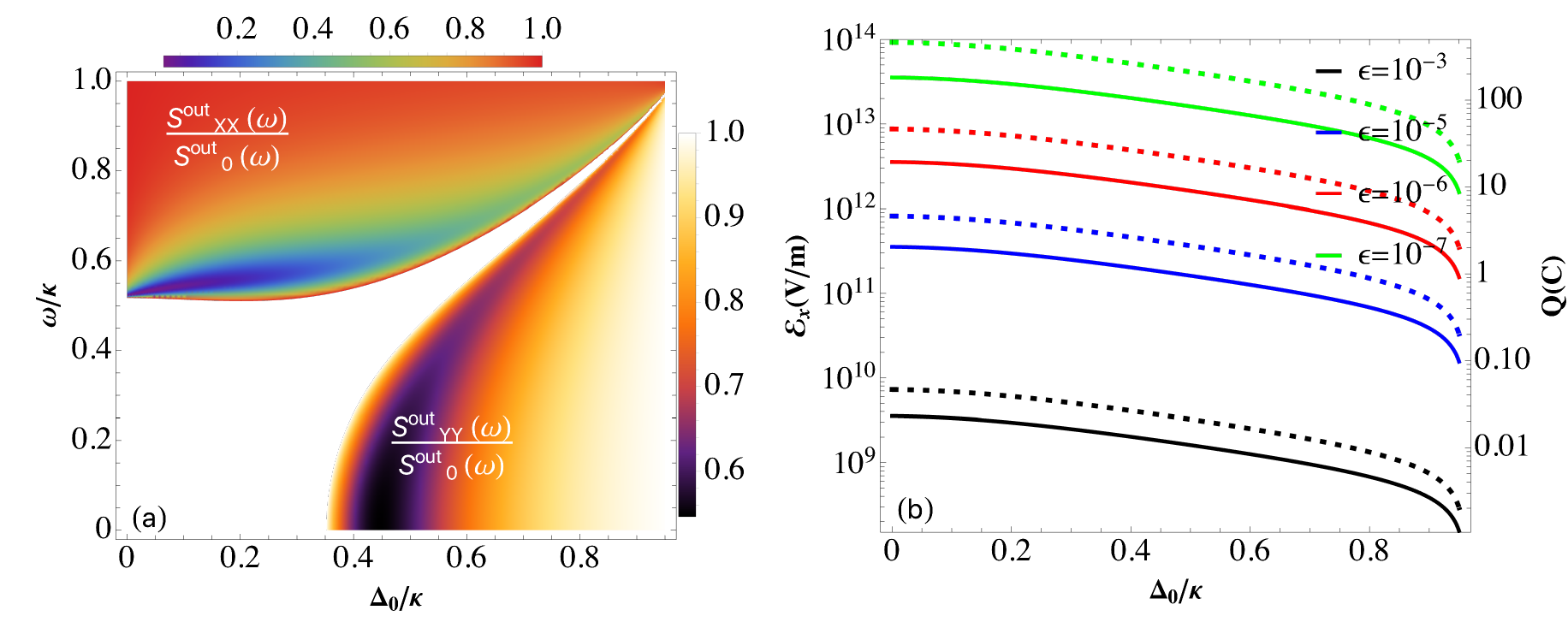}
\caption{(color online) (a) Density plot of the symmetric spectral function $S^{\rm{out}}_{\rm{JJ}}(\omega)/S^{\rm{out}}_{0}(\omega)$ for the amplitude (J=X, above the white region) and phase (J=Y, below the white region) quadrature fluctuations of the output light against $\omega/\kappa$ and $\Delta_{0}/\kappa$, where the white region indicates the absence of squeezing. (b) The plot displays the electrostatic field ${\cal E}_{\rm{x}}$ (solid curves) in V/m and the charge on the ring $Q $ (dashed curves) in  $C$  as functions of $\Delta_{0}/\kappa$ for various values of mCP charge q with the parameter $\epsilon$, including $10^{-3}$ (black curve), $10^{-5}$ (blue curve), $10^{-6}$ (red curve) and $10^{-7}$ (grean curves). All other parameters remain consistent with Fig. 1 of the main text.} \label{sp1}
\end{figure}
\begin{figure} 
\includegraphics[scale=.55]{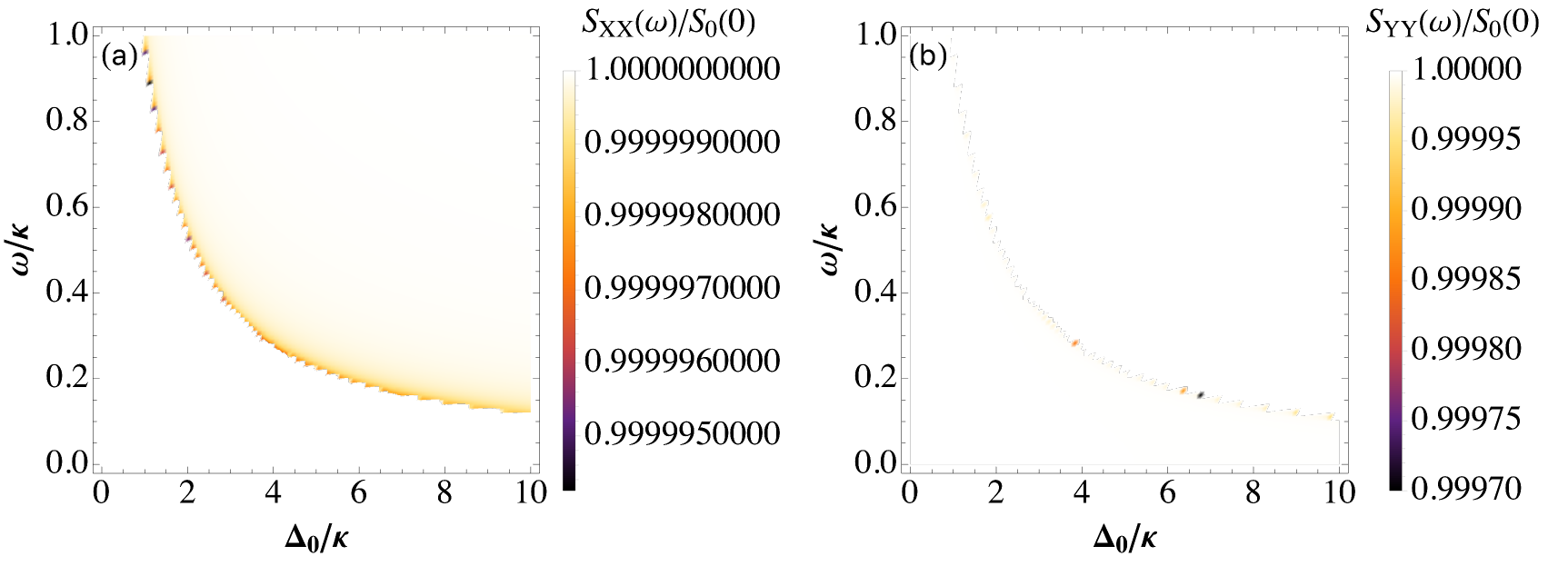}
\includegraphics[scale=.55]{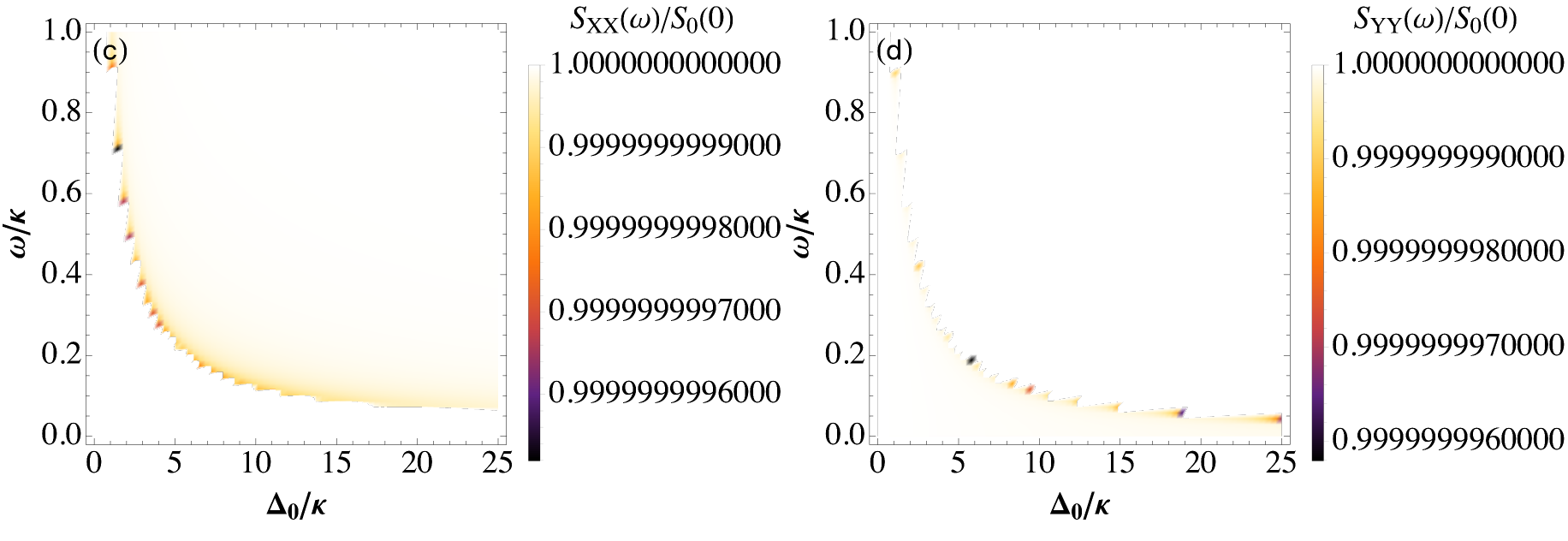}
\caption{(color online) Density plot of the symmetric spectral function, normalized to the thermal spectrum, of the output amplitude quadrature fluctuations $S_{\rm{XX}}^{\rm{out}}(\omega)/S^{\rm{out}}_{0}(\omega)$ (a, c) and phase quadrature $S_{\rm{YY}}^{\rm{out}}(\omega)/S^{\rm{out}}_{0}(\omega)$ (b ,d) is presented as a function of $\omega/\kappa$ and normalized detuning $\Delta_0/\kappa$ at room temperature for $\epsilon= 10^{-8} $ (a,b), $\epsilon= 10^{-10} $ (c,d) , and a fixed value of the electrostatic field $\mathcal{E}_{\rm{x}}=7.25\times 10^{10}$V/m. Other parameter values are provided in the main text, illustrating the existence of squeezing in the output light for these quadratures at slightly different values of $\omega/\kappa$.}\label{sp2}
\end{figure}

In usual opto-mechanical systems to observe quantum effects like entanglement between the mechanical mode and the internal cavity light the effective detuning is fixed at the Stokes or anti-Stokes sideband. Here the Stokes will be chosen because at the anti-Stokes side band the drift matrix $\bf A$ has non negative eigenvalues real parts and system becomes unstable. This choice introduces another constraint. Since any interesting quantity depends on $\Delta_{0}$,  one has to satisfy the following transcendental system
\begin{eqnarray} \label{sys}
\frac{4\hbar g k\sin(2kx_{\rm{s}})|E|^2}{\kappa^2+4[\Delta_{0}+g\cos^2(kx_{\rm{s}})]^2}& =&-A_{\rm{q}}(x_{\rm{s}}+C_0) \label{sys1} , \\
\frac{8\hbar g k^2 |E|^2\cos(2kx_{\rm{s}})}{\kappa^2 +4 [\Delta_{0} + g\cos^2(k x_{\rm{s}})]^2}&=&m[\Delta_{0} + g \cos^2(k x_{\rm{s}})]^2. \label{sys2}
\end{eqnarray}
The dynamics of the system is influenced by different parameters, each of which plays a significant role in determining the value of $x_{\rm{s}}$. When the input power ${\cal P}$ is fixed, the optimal approach to assess the presence of entanglement and squeezing is as follows: first, set $\Delta_{0}$ and solve Eq. (\ref{sys2}) to determine $x_{\rm{s}}$.  Subsequently, usnig Eq. (\ref{sys1}) to find the charge value $Q$ on the metallic ring, considering a specific mCP charge and displacement $C_0$. Only those combinations of parameters $a_{\rm{s}}$, $\omega_{\rm{m}}$, $\Omega_{\rm{m}}$, and $G$ that ensure system stability are taken into account. 

Hence we analyze this situation in Figs.\ref{sp1}(a) where symmetric spectral function, normalized to the thermal spectrum, of the output amplitude quadrature $S_{\rm{XX}}^{\rm{out}}(\omega)/S^{\rm{out}}_{0}(\omega)$ and phase $S_{\rm{YY}}^{\rm{out}}(\omega)/S^{\rm{out}}_{0}(\omega)$  quadrature fluctuations as a function of normalized detuning $\Delta_{0}/\kappa$ and $\omega/\kappa$ for any chosen mCP charge $q$, with the parameter $\epsilon$ raging from $10^{-3}$ to $10^{-7}$. The corresponding optimal values of the ring charge $Q$ (represented by the solid curves) and, consequently, the electrostatic field at position $x_{\rm{s}}$: ${\cal E}_{\rm{x}}(x_{\rm{s}})\simeq {Q (C_0 + x_{\rm{s}})}/(4\pi \epsilon_0 R^3)$ (illustrated by the dashed curves) as function $\Delta_{0}/\kappa$ are shown in Fig. \ref{sp1}(b). Significant squeezing is observed in both the amplitude and phase quadratures of the output field by choosing appropriate values for the electrostatic field ${\cal E}_{\rm{x}}(x_{\rm{s}})$ and the ring charge $Q$, corresponding to different values of the mCP charge $q$, as shown in Fig.\ref{sp1}(b).For a small value of mCP charge,  Fig.\ref{sp1}(b) shows that a strong electrostatic field ${\cal E}_{\rm{x}}(x_{\rm{s}})$ and high corresponding charge $Q$ are actually required for squeezing in the output quadrature of field. 

Figure \ref{sp2} shows, instead, the results when $\Delta(x_{\rm{s}})\neq \omega_{\rm{m}}$ for a fixed value of electrostatic field $\mathcal{E}_{\rm{x}}(x)$. 
In this case, it is also possible to observe a squeezing effect in the output field's amplitude and phase quadrature fluctuations. For example, in the case of an electrostatic field strength of ${\cal E}_{\rm{x}} = 2.25 \times 10^{10}$ V/m, the results in Fig.\ref{sp2}, would correspond to different values of mCP charge $q$. Specifically first row (Figs. \ref{sp2}(a-b)) is evaluated for $\epsilon = 10^{-8}$ while the second row (Figs. \ref{sp2}(c-d)) of Figs. \ref{sp2} corresponds to $\epsilon$=$10^{-10}$. Despite the small interaction between mCP and our world, according to the parameters chosen, the output light from the cavity is no longer thermal noise, but rather squeezed light.

The output spectra of the phase and amplitude quadratures show a noticeable sensitivity to the correlations of the noises. Both spectra show values below the thermal noise threshold for a range of $\omega/\kappa$ values, which varies for amplitude and phase quadrature variations, as shown in Figs.\ref{sp1} and \ref{sp2}. This implies that at these frequencies, the output light is subject to amplitude and phase quadrature squeezing. Notably, the squeezing effect is a quantum phenomenon caused by the presence of mCP particles. This field's potency is further influenced by the displacement $C_0$ of the ring plane. This intriguing scenario could potentially herald another consequential quantum effect: entanglement. However, in the case of entanglement, one is limited to considering mCP charges $q \ge 10^{-6} e_0$; otherwise, a very strong electrostatic field is required, as depicted in Fig.\ref{sp1}(b). 

\section{Entanglement and Elements of Co-variance matrix}
Finally, to quantify the entanglement, the value of the logarithmic negativity \cite{S_VidalWerner, S_Plenio05, S_adesso} is considered, given by
\begin{equation}\label{ent}
E_n = \max\{0, -\log(2\eta_-)\},
\end{equation}
where $\eta_-$ represents the lowest symplectic eigenvalue of the partially transposed correlation matrix $\mathbf{V}$, obtained as \cite{S_simon}
\begin{equation} \label{eigen}
\eta_{-} = \sqrt{\frac{\sigma - \sqrt{\sigma^2 - 4\det{\mathbf{V}}}}{2}},
\end{equation}
with
%\begin{equation}
$\sigma = \det{\mathbf{B_1}} + \det{\mathbf{B_2}} - 2\det{\mathbf{B_3}},$
%\end{equation}
and the $2 \times 2$ sub-matrices $\mathbf{B_1}$, $\mathbf{B_2}$, and $\mathbf{B_3}$ in terms of the elements of the correlation matrix $\mathbf{V}$ are given by
\begin{equation}
\mathbf{B_1} = \begin{pmatrix} V_{11} & V_{12}  \\ V_{21} & V_{22} \end{pmatrix}, \quad
\mathbf{B_2} = \begin{pmatrix} V_{33} & V_{34} \\ V_{43} & V_{44} \end{pmatrix}, \quad
\mathbf{B_3} = \begin{pmatrix} V_{13} & V_{14} \\ V_{23} & V_{24} \end{pmatrix}.
\end{equation}
In linearized approximation, both modes—the internal cavity mode and NS' CoM—are represented by Gaussian states due to the delta-correlated nature of the noises in time and the linearity of the dynamics. As a result, all relevant information is extracted from the correlation matrix $\mathbf{V}$, whose elements are defined by $V_{\rm{ij}}=[\langle u_{\rm{i}}(\infty)u_{\rm{j}}(\infty)+u_{\rm{j}}(\infty)u_{\rm{i}}(\infty)\rangle]/2$. These elements are obtained by solving the following Lyapunov equation \cite{S_vitalietal}:
\begin{equation} \label{lya}
\mathbf{A} \cdot \mathbf{V} + \mathbf{V} \cdot \mathbf{A}^T = -\mathbf{D}.
\end{equation}
Here, the diffusion matrix $\mathbf{D} = \text{diag}(0, \Gamma, \kappa/2, \kappa/2)$ serves as a diagonal matrix that accounts for noise correlations, as defined in the main text.  The $4 \times 4$ symmetric matrix {\bf V} can be written as 
\begin{equation}
{\bf V} = {\bf V_0} + G^2 {\bf F},
 \end{equation}
where the diagonal matrix ${\bf V_0}$ is  
\begin{equation}
{\bf V_0}=\text{diag}\left(\frac{\Gamma \omega_{\rm{m}} }{\gamma \Omega_{\rm{m}}},\frac{\Gamma}{\gamma},1/2,1/2\right), 
\end{equation}
and ${\bf F}$ is a $4 \times 4$ symmetric matrix with intricate expressions that are not reported here. Therefore, it can be immediately concluded that for $x_{\rm{s}}=0$, $E_n =0$. 


\begin{thebibliography}{51}
\bibitem [{sup({\natexlab{a}})}]{supmatb}%
  \BibitemOpen
  \href@noop {} {}\bibinfo {howpublished} {However, similar results to those
  described here could be obtained if a few electrons are left on the surface
  of the silica droplet. In that case, no dark matter could be observed, but
  interesting quantum effects may still arise} ({\natexlab{a}})\BibitemShut
  {NoStop}%
\end{thebibliography}
\end{document}